# Comparison of $CO_2$ trapping in highly heterogeneous reservoirs with Brooks-Corey and van Genuchten type capillary pressure curves

Naum I. Gershenzon1[a], Robert W. Ritzi Jr.[a], David F. Dominic[a], Edward Mehnert[b],

Roland T. Okwen[b]

[a]Department of Earth and Environmental Sciences, Wright State University, 3640 Col. Glenn Hwy., Dayton, OH 45435, USA

[b]Illinois State Geological Survey, Prairie Research Institute, University of Illinois at Urbana-Champaign, 615 East Peabody Drive, Champaign, IL 6182, USA

## Abstract

Geological heterogeneities essentially affect the dynamics of a $CO_2$ plume in subsurface environments. Previously we showed how the dynamics of a $CO_2$ plume is influenced by the multi-scale stratal architecture in deep saline reservoirs. The results strongly suggest that representing small-scale features is critical to understanding capillary trapping processes. Here we present the result of simulation of $CO_2$ trapping using two different conventional approaches, i.e. Brooks-Corey and van Genuchten, for the capillary pressure curves. We showed that capillary trapping and dissolution rates are very different for the Brooks-Corey and van Genuchten approaches when heterogeneity and hysteresis are both represented.

* Corresponding author. Tel.: +1-937-775-2052.
  *E-mail address:* naum.gershenzon@wright.edu

# 1. Introduction

There are various trapping mechanisms of supercritical $CO_2$ in saline reservoirs ranging from short term (0 – 10 years) capillary trapping to middle term (0 - 1000 years) dissolution in brine to long term (100 years – million years) mineralization. Here we will consider only short term processes associated with capillary trapping and dissolution.

The cause of $CO_2$ trapping in rock pores (in gas phase) is the difference between capillary pressure of brine and supercritical $CO_2$, resulting in disconnection of the $CO_2$ stream into immobile blobs and ganglia [1]. Here and below the terms "gas phase" and "gas saturation" are short for "supercritical gas phase" and "supercritical gas saturation". Capillary trapping in homogeneous porous media incorporates three related mechanisms, i.e. residual trapping, trapping due to hysteresis of the relative permeability, and trapping due to hysteresis of the capillary pressure. Kumar et al. [2], analysing residual and aqueous $CO_2$ trapping, showed that the former is a dominant short-term trapping mechanism. Juanes et al. [3] found that relative permeability hysteresis becomes a major factor in $CO_2$ trapping. Altundas et al. [4] showed that not only hysteresis of permeability but also hysteresis of capillary pressure affects the dynamic of $CO_2$ trapping. The latter mechanism works not only on the tail of the $CO_2$ plume but at front as well, causing its retardation. The conventional approach for modelling of $CO_2$ plume dynamics involves capillary pressure and relative permeability curves, i.e. function or table of capillary pressure and relative permeability of brine and $CO_2$ versus brine saturation for drainage and imbibition. The shape and particular characteristics of these curves, such as irreducible water saturation, $S_{wi}$, maximum residual gas saturation (trapped saturation), $S_{gr}$ and the entry pressure point into the pore of the rock, $P_e$, play a key role in the $CO_2$ trapping processes (both by capillary trapping and dissolution) and define the rate and amount of trapped $CO_2$ as well as the shape and dynamics of the $CO_2$ plume [2-11]. Two common capillary pressure curves are the Brooks-Corey [12] or the van Genuchten [13] models. Hereafter we will use abbreviations BC and vG to designate Brook-Corey and van-Genuchten-type capillary pressure curves. Li et al. [11] investigated the difference between simulations of $CO_2$ sequestration in aquifers using these two models. They observed that 1) the vG-type capillary pressure model accelerates $CO_2$ solubility trapping compared with the BC-type model; 2) simulation results are very sensitive to the slope of the vG curve around the entry-pressure region.

Geological heterogeneities essentially affect the dynamics of a $CO_2$ plume in subsurface environments [5, 8, 14-20]. Hovorka et al. [14] concluded that stratigraphic heterogeneity may result in significant channelling of the flow of a $CO_2$ plume. Doughty and Pruess [5] created three-dimensional models representing the fluvial/deltaic Frio Formations in the upper Texas Gulf Coast. Their simulations also demonstrate the strong influence of geologic heterogeneity on buoyancy-driven $CO_2$ migration. Obi and Blunt [15], considering a lateral movement (advection) of the $CO_2$ plume under a horizontal pressure gradient, found that permeability heterogeneity causes the $CO_2$ to migrate much farther than predicted from a homogeneous model. Bryant et al. [16] and Ide et al. [8] showed that a layer of smaller permeability lying above a layer of larger permeability can act as a seal due to capillary pressure barrier. Saadatpoor et al. [17] further exposed this additional trapping mechanism, which they termed a "local trapping mechanism", associated with the heterogeneity of capillary pressure. Supposing a functional dependence between permeability and capillary pressure, they simulated the dynamics of the buoyant $CO_2$-brine front in a heterogeneous reservoir. Comparison of this simulation with an analogous simulation but with a homogeneous (averaged) capillary pressure curve revealed a dramatic difference in results. In the former case $CO_2$ raises through the high permeability channels, which are surrounded by the capillary barriers of the low permeability material. In some regions capillary barriers prevent upward movement of $CO_2$, allowing only lateral migration, which effectively traps the $CO_2$ plume. The same idea has been exploited by Zhou et al. [18]. Considering regional-scale (typical lateral size up to a few km) flow and transport processes in a layered reservoir, they investigated the so-called "secondary-seal effect", i.e. the effect of $CO_2$ accumulation between layers with different permeability and capillary pressure entry points, causing retardation of upward $CO_2$ migration.

Recent studies have led to new conceptual and quantitative models for sedimentary architecture in fluvial deposits over a range of scales that are relevant to the performance of some deep saline reservoirs [21, 22]. Previously, we showed how the dynamics of a $CO_2$ plume, during and after injection, is influenced by the hierarchical and multi-scale stratal architecture in such reservoirs [23, 24]. The results strongly suggest that representing these small-scale (few cm in vertical direction and few meters in horizontal direction) features and representing how they are organized within a hierarchy of larger-scale features, is critical to understanding capillary trapping processes. The results also demonstrated the importance of using separate capillary pressure and relative permeability relationships for different textural facies. Here we compare the result of simulation of $CO_2$ trapping in deep saline aquifers using two different conventional approaches to capillary pressure – $CO_2$ saturation curves, i.e. BC and vG. We show that capillary trapping as well as dissolution rates are very different for the BC and vG approaches if reservoir consists from various species with different capillary pressure and relative permeability curves.

## 2. Methodology

The methodology of these simulations has been described in detail in Gershenzon et al. [23, 24]. Here we use the same reservoir geometry and all other parameters including capillary pressure and relative permeability curves. The only addition is the vG-type capillary pressure curves (see Fig. 1). This methodology uses a code developed by Ramanathan et al. [25] and Guin et al. [26] to generate a realization of a heterogeneous reservoir including two materials, i.e. sandstone (76%) and open framework conglomerate (OFC) (24%) with geometric mean permeability 58 mD and 3823 mD, respectively. The reservoir size is 100 m x 100 m x 5 m (250 thousand cells of size 2 m x 2 m x 0.05 m). $CO_2$ was injected at a rate of 3.6 (standard) m³/day during 10 days into the bottom of a vertical well at a depth 2360 m. The well was placed at x = 23 m y = 1 m. No flow boundaries were imposed on all boundaries of the $CO_2$ reservoir. The total amount of injected $CO_2$ is 594 kg. Two different sets of characteristic curves were utilized for the sandstone and OFC unit types. Thus, the total number of characteristic curves was 12 including six for drainage and six for imbibition (see [24] for more details). To generate BC-type capillary pressure $P_c$ curve for drainage we use the relation:

$$P_c = P_e \left(\frac{S - S_{wi}}{1 - S_{wi}}\right)^{-1/\lambda} \tag{1}$$

where $S$ is brine saturation, $P_e$ is the minimum pressure required for the entry of $CO_2$ into the pore of the rock, $S_{wi}$ is irreducible water saturation, and $\lambda$ is a fitting parameter known as pore size distribution index. To generate a vG-type capillary pressure curve we follow [11] and use the same relation (1) in the range $S_{wi} \leq S \leq (1 - S_{nt})$ ($S_{nt}$ is the width of the entry slope region):

$$P_c = (1-S)\frac{P_e}{S_{nt}}\left(\frac{S - S_{nt} - S_{wi}}{1 - S_{wi}}\right)^{-1/\lambda}, \text{ if } 1 - S_{nt} < S \leq 1 \tag{2}$$

The parameter values used in our simulation are $\lambda = 0.55$, $P_e(\text{sandstone}) = 4.6 \cdot 10^{-2}$ bar,

$P_e$(OFC) $= 0.72 \cdot 10^{-2}$ bar, $S_{wi}$(sandstone) $= 0.22$, $S_{wi}$(OFC) $= 0.143$, and $S_{nt} = 0.1$. The imbibition curves are the same as in Gershenzon et al. [23, 24]. There is no difference between BC and vG types for imbibition.

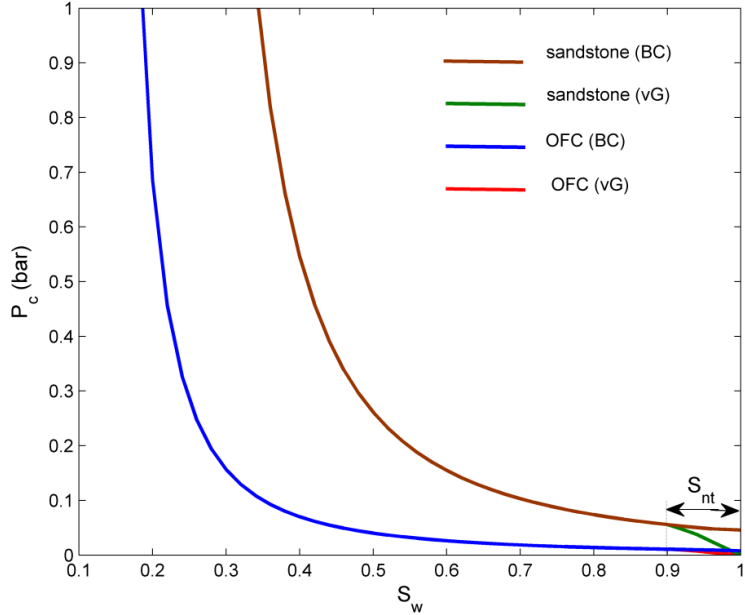

Fig. 1. Brooks-Corey and van Genuchten capillary pressure curves for OFC and sandstone for drainage.

## 3. Results

Fig. 2 depicts the dynamics of $CO_2$ trapping of the gas phase in homogeneous and heterogeneous reservoirs in cases of simulation with BC and vG types of capillary pressure curves. Note that in homogeneous reservoirs capillary trapping is virtually not affected by the type of capillary pressure – $CO_2$ saturation relation (compare red and green curves in Fig. 2) as was already noticed by Li et al. [11]. This is because capillary trapping depends mostly on differences in the shape and parameters of the relative permeability curves for drainage and imbibition between facies types. However, in the heterogeneous case it is no longer true, as is clearly demonstrated in Fig. 2 (compare aqua and blue curves). Capillary trapping is about 12% larger in the vG case. The difference is even more dramatic in each facies. Indeed, trapping in sandstone is more than 60% larger in the vG case (compare aqua and green curves in Fig. 3). In contrast, trapping in OFC material is about 20% smaller in the vG case (compare red and blue curves in Fig. 3). To explain these effects let us recall that in heterogeneous reservoirs (which include two or more different facies with respectively different values of the capillary pressure) the gradient of pressure and/or buoyancy force, which push $CO_2$ through the facies boundary (say from OFC to sandstone), should overcome capillary pressure. In contrast, the capillary pressure pushes off $CO_2$ from sandstone to OFC. As a result, part of the $CO_2$ is accumulated at the boundaries between facies (secondary-seal effect). In this case the $CO_2$–brine interface zone coincides with OFC–sandstone boundary. The difference between BC and vG representation is reflected in the width of $CO_2$–brine interface zone. In the latter case the width is larger due to a small part of the $CO_2$ crossing the OFC–sandstone boundary without an entry pressure obstacle.

This difference explains 1) why sandstone traps more $CO_2$ and OFC traps less $CO_2$ in vG case than in BC case and 2) why the total amount of capillary trapping in an entire reservoir is larger in vG case (see Discussion and Conclusions section for more detail explanation).

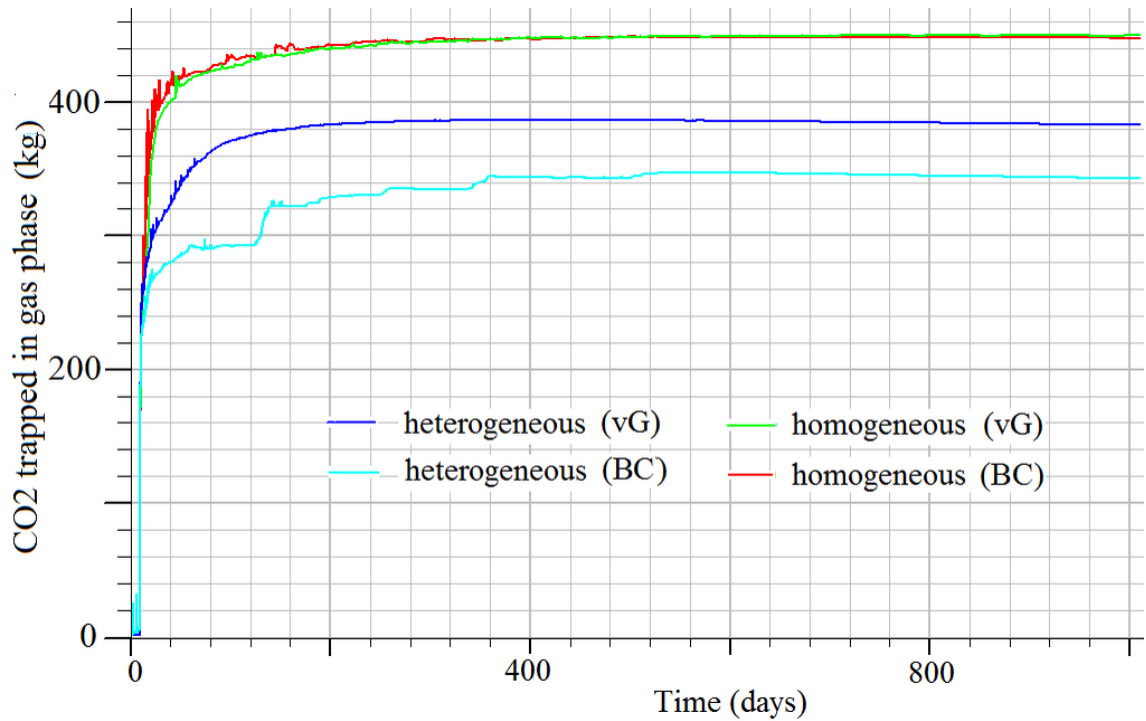

Fig. 2. The dynamics of $CO_2$ trapping in gas phase in homogeneous and heterogeneous reservoirs in case of simulation with BC and vG capillary pressure curves.

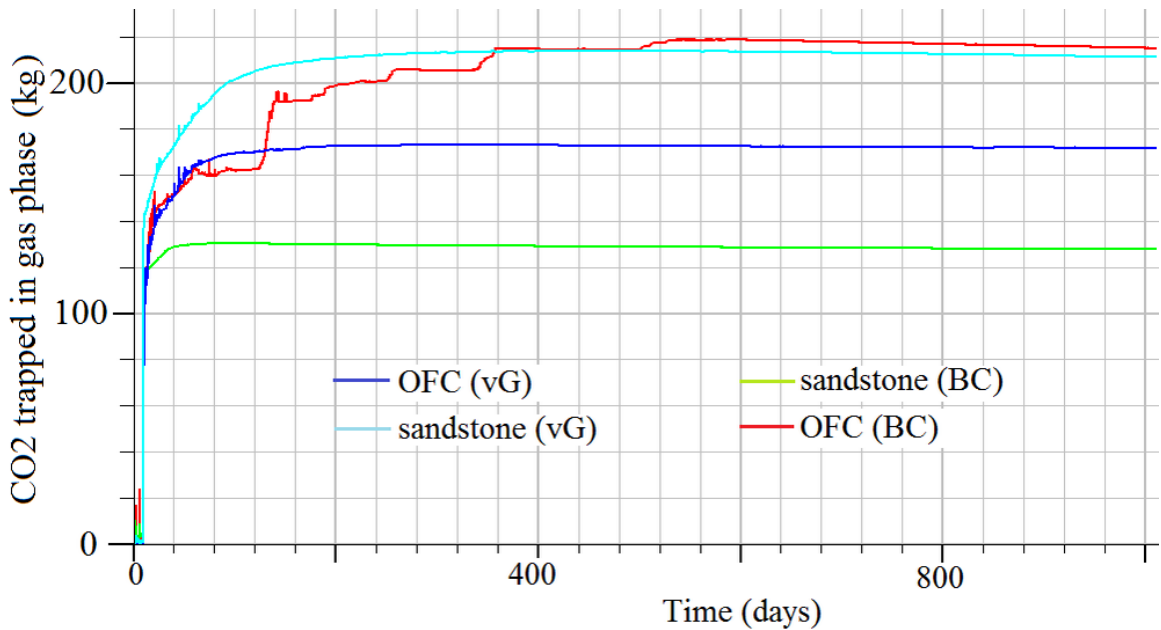

Fig. 3. The dynamics of $CO_2$ trapping in gas phase (in heterogeneous reservoir) for OFC and sandstone for BC and vG capillary pressure curves.

Fig. 4 shows the dynamics of $CO_2$ dissolution in homogeneous and heterogeneous reservoirs for the simulations with BC and vG-type capillary pressure curve. Notice the difference (about 20%) between dissolved $CO_2$ in homogeneous reservoir simulated using BC and vG curves (red and green curves in Fig. 4). This effect, also described by Li et al [11], can be explained by the difference between effective widths of the brine- $CO_2$ interface zone.

In the heterogeneous reservoir the amount of dissolved $CO_2$ (simulated with vG curve) is almost double compared with homogeneous reservoir (green and blue curves in Fig. 4). This is not surprising since the $CO_2$-brine interface zone area is much larger in heterogeneous case (compare plume shape at the top and the bottom panels in Fig. 2 in Gershenzon et al. [23]). The difference between dissolved $CO_2$ in the heterogeneous reservoir simulated using BC and vG curves (aqua and blue curves in Fig. 4) is even more dramatic (30-100%) than in homogeneous case due to the already described difference between widths of the brine- $CO_2$ interface zone adjusted to the OFC-sandstone boundaries. Thus, the effect of dissolution increase due to vG approach is multiplied in highly heterogeneous reservoirs.

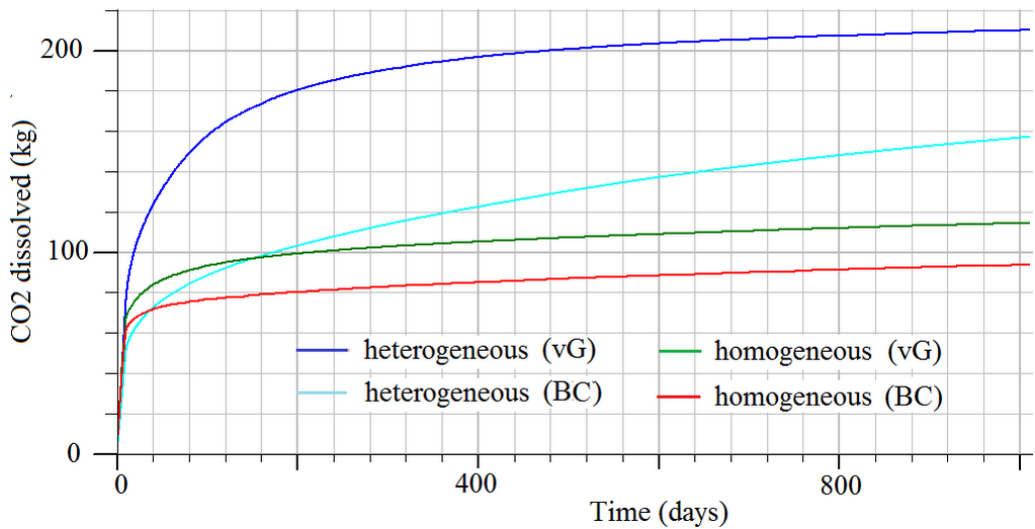

Fig. 4. The dynamics of $CO_2$ dissolved in homogeneous and heterogeneous reservoirs in case of simulation with BC and vG capillary pressure curves.

The difference between mobile $CO_2$ in the gas phase for the BC and vG cases is dramatic for the heterogeneous reservoir, especially at the end of the process (see Fig. 5). For the simulation conditions considered, the amount of mobile $CO_2$ is negligible for the vG case. By the end of the simulation the ratio between amounts of mobile $CO_2$ for BC and vG cases is more than 50. This is the result of rapid dissolution and capillary trapping of $CO_2$ in vG case. Note that the amount of the mobile $CO_2$ in the heterogeneous reservoir is 1.5 - 3 times larger than in the homogeneous reservoir in BC case. However, as we showed previously [23, 24], the remaining gas (not dissolved and not capillary trapped) in the heterogeneous reservoir in this particular example is actually trapped due to the secondary-seal effect and placed below OFC – sandstone boundaries. This is also seen in the Fig. 6 which depicts the spatial distribution of $CO_2$ saturation at the end of simulation for the BC and vG cases. As seen in Fig. 6, the $CO_2$ plumes never reach the top of the reservoir.

From the Fig. 6 we can also see that the shape of the plume is basically the same in BC and vG cases since this shape is defined by the shape of the OFC clusters and by their positions relative to the injection well. The small difference between shapes and the amount of $CO_2$ in gas phase (seen as difference in plume color) are due to difference in amount of dissolved $CO_2$.

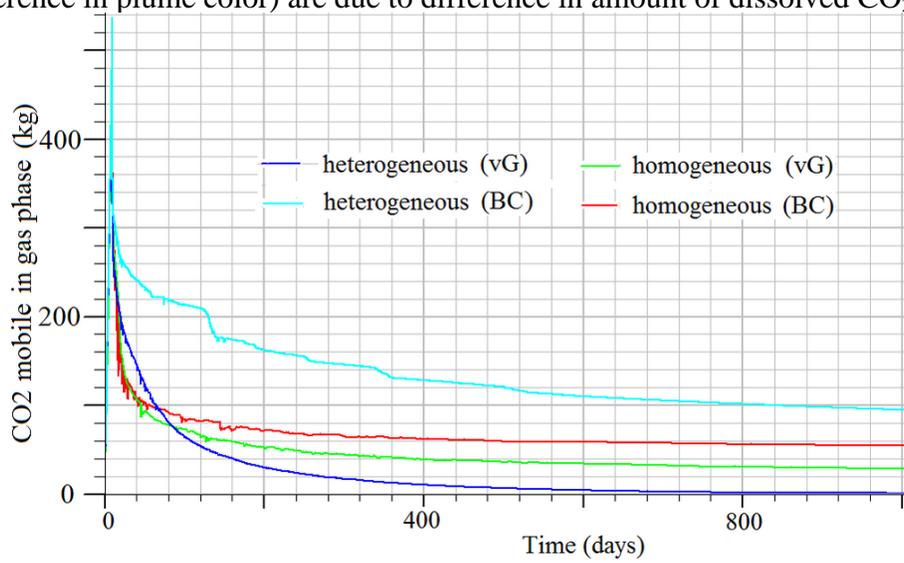

Fig. 5. The dynamics of mobile $CO_2$ in the gas phase in homogeneous and heterogeneous reservoirs in simulations with BC and vG capillary pressure curves.

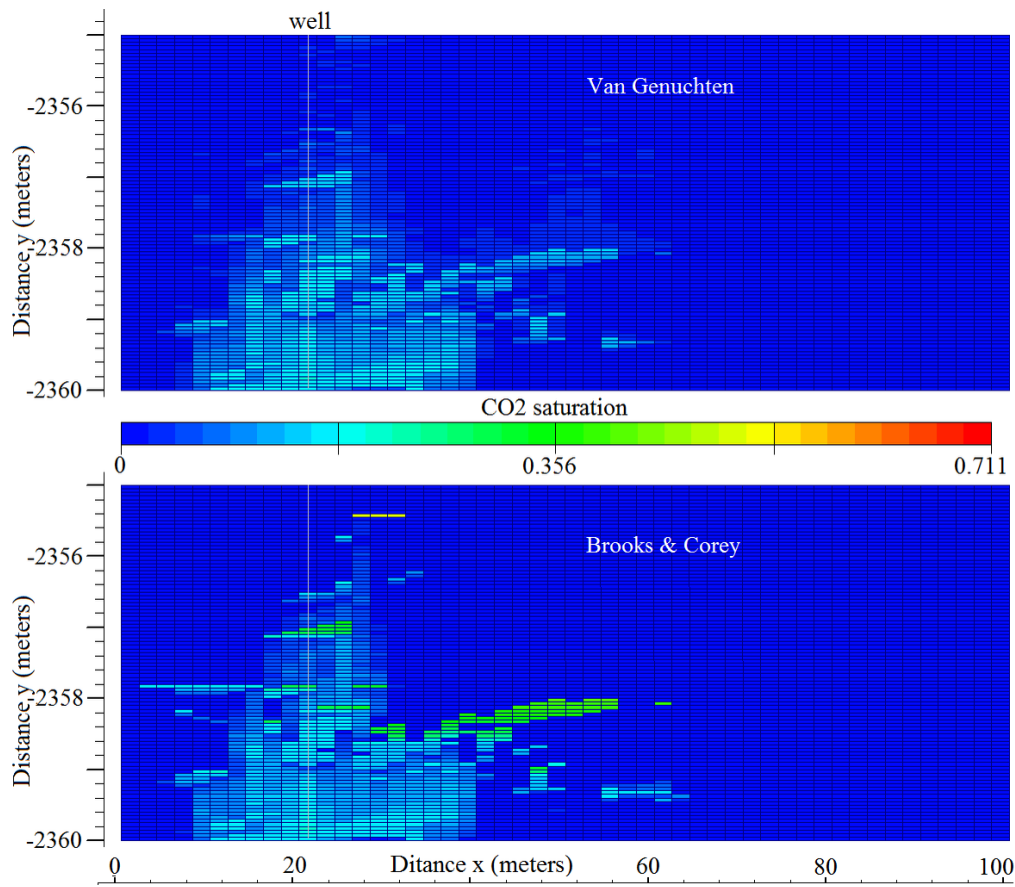

Fig. 6. Vertical cross section of reservoir showing $CO_2$ saturation after 1010 days (1000 days after the end of injection) in case of vG (top panels) and BC (bottom panel).

To study the influence of the size of $S_{nt}$ (the "vG tail") we made a simulation with $S_{nt}$=0.04, i.e. 2.5 times smaller than in the previous simulation. The comparison of simulation results between case vG1 ($S_{nt}$=0.1), case vG2 ($S_{nt}$=0.04) and case BC shows that 1) the amount of mobile $CO_2$ in gas phase in case vG2 is a little bit larger than in case vG1 case but still much smaller than in case BC; 2) the amount of trapped $CO_2$ in gas phase in case vG2 is smaller than in case vG1 case but still larger than in case BC; 3) the amount of dissolved $CO_2$ in case vG2 is a little bit larger than in case vG1 case and much larger than in case BC.

## 4. Discussion and Conclusions

Simulation of $CO_2$ sequestration requires knowledge of capillary pressure – $CO_2$ saturation relations. These functional dependences are difficult to derive experimentally and they are quite uncertain for most potential reservoirs. Conventionally such curves are derived from a few experimentally defined points using Brooks & Corey [12] or van Genuchten [13] approaches. The difference between these two approaches is in the form of the capillary pressure curve where $CO_2$ saturation is small. These results show that even small changes in the capillary pressure curve at low $CO_2$ saturation visibly affect the rate and the amount of $CO_2$ trapping and therefore distinguishing the true capillary pressure at small $CO_2$ saturation may be important.

Li et al. [11] found that in homogeneous reservoirs $CO_2$ dissolution was affected by the shape of the capillary pressure tail, but capillary trapping was insensitive to this shape. Here, we investigated the difference in $CO_2$ trapping (both dissolution and capillary trapping) between BC and vG types of capillary pressure curves for a heterogeneous reservoir including two types of rock (sandstone and open framework conglomerate). We found:

1. Capillary trapping is significantly impacted by the type of capillary pressure curve used. Capillary trapping averaged over the whole reservoir is larger in the vG case. This effect is even larger for the finer-grained sandstone, but is negative for the coarser-grained OFC.

2. $CO_2$ dissolution is considerably larger in case of the heterogeneous reservoir than in the homogeneous reservoir when capillary pressure is defined using a vG type curve.

The rate of dissolution in a homogeneous reservoir is proportional to the area of the plume surface. The effective surface area is larger for the $CO_2$-brine interface with larger width. This explains the effect of increase of dissolution in the vG case. The same explanation is applicable for the heterogeneous reservoirs. However, the boundaries between different material (OFC and sandstone in our case) should be considered. In heterogeneous media $CO_2$ propagates along high permeable clusters in addition to moving up by buoyancy force. As a result, a part of the $CO_2$ (in our example all $CO_2$ in gas phase) is trapped at the OFC – sandstone boundaries (secondary-seal effect) (see Fig. 6), because the buoyancy force is smaller than the capillary pressure force. The latter force prevents $CO_2$ moving upward. The penetration width of $CO_2$ at the OFC – sandstone boundary depends on the pore size distribution in sandstone. The wider the distribution, the larger the effective $CO_2$ – brine surface, and the larger the dissolution rate, which explains the results depicted at Fig. 4. This effect is much larger in heterogeneous reservoirs than in homogeneous reservoirs since the total effective contact area is much larger.

Capillary trapping in homogeneous media does not depend on the width of $CO_2$ – brine interface zone, and hence does not depend on the type of capillary pressure curve used for simulation. That is why in homogeneous reservoirs the capillary trapping rate and the amount are the same for BC and vG curves (see Fig. 2). In heterogeneous reservoirs the $CO_2$ – brine interface zone partially (or entirely as in our example) coincides with OFC – sandstone boundaries (see Fig. 6). In this case the width of the $CO_2$ – brine interface zone affects the distribution of $CO_2$ between OFC and sandstone. In the BC case the width of the interface zone is smaller and $CO_2$ does not penetrate up into sandstone. In contrast, in the vG case a part of the $CO_2$ penetrates into sandstone. As a result, the ratio between amounts of $CO_2$ in sandstone to amounts of $CO_2$ in OFC is larger in the vG case. This explains why the amount of capillary trapped $CO_2$ is larger in sandstone (and respectively is smaller in OFC) in the vG case than in the BC case (Fig. 3). Since the capillary trapping is larger in sandstone than in OFC (per unit volume of material) the total amount of capillary trapped $CO_2$ in the entire (heterogeneous) reservoir is larger in the vG case (Fig. 2).

## Acknowledgements

This work was supported as part of the Center for Geological Storage of $CO_2$, an Energy Frontier Research Center funded by the U.S. Department of Energy, Office of Science, Basic Energy Sciences under Award # DE-SC0C12504. We acknowledge Schlumberger Limited for the donation of ECLIPSE Reservoir Simulation Software. This work was supported in part by the Ohio Supercomputer Center, which provided an allocation of computing time and technical support.